\documentclass[journal, table]{IEEEtran}

\usepackage{amsmath, amssymb, booktabs, bm, cite, color, lipsum, siunitx, mathdots, multirow, textcomp, verbatim, xcolor, cuted, stfloats}
\usepackage{threeparttable}
\usepackage[section]{placeins}
\allowdisplaybreaks[3]
\usepackage{longtable}
\usepackage{graphicx}
\graphicspath{{figures/}}

\ifCLASSOPTIONcompsoc
  \usepackage[caption=false,font=normalsize,labelfont=sf,textfont=sf]{subfig}
\else
  \usepackage[caption=false,font=footnotesize]{subfig}
\fi

\usepackage{mdwmath}
\usepackage{eqparbox}
\usepackage{url}
\usepackage{algorithm}
\usepackage{algorithmicx}
\usepackage{algpseudocode}

\usepackage{booktabs} 
\usepackage{multirow} 
\usepackage{lipsum}   
\usepackage{tikz}
\usepackage{siunitx}
\usepackage{tabularx}
\usepackage{makecell}
\usepackage{booktabs}
\usepackage{multirow}

\usetikzlibrary{shapes,arrows,fit,positioning}

\usepackage{etoolbox}
\usepackage{hyperref}
\newtoggle{hl}
\togglefalse{hl}
\iftoggle{hl}{%

}{%

}

\begin{document}
\pagestyle{empty}
\bstctlcite{IEEEexample:BSTcontrol}
    \title{Unified Error Analysis of Multi-site Radar via Equivalent Angular Resolution}
\author{%
\IEEEauthorblockN{%
Lang Qin\textsuperscript{\#\$}, 
Zelin Liu\textsuperscript{\$}, 
Rongjie Li\textsuperscript{\$},
Zhiqiang Huang\textsuperscript{\#},
Xiaoguang Liu\textsuperscript{\$}\\
}
\IEEEauthorblockA{%
\textsuperscript{\#}The Hong Kong University of Science and Technology (Guangzhou), China\\
\textsuperscript{\$}Southern University of Science and Technology, China\\
}
}


\maketitle

\thispagestyle{empty} 
\begin{abstract}

High-precision indoor sensing using monostatic multiple-input multiple-output (MIMO) radar typically relies on increasing the physical aperture size of antennas, leading to high hardware complexity and cost. To overcome this bottleneck, this paper establishes a unified framework for multi-site radar sensing based on equivalent angular resolution, together with a design methodology that uses this metric to optimize distributed Single-Input Single-Output (SISO) configurations. By mapping spatial diversity into the angular domain, the proposed metric enables a direct and physically interpretable comparison with monostatic MIMO beamwidth. The associated methodology provides a principled way to select node placement and geometry to synthesize an effective virtual aperture that suppresses angular glint and multipath. Experiments with commercial 60-GHz radars in cluttered indoor environments validate the superiority of the multi-site SISO configuration over monostatic MIMO, demonstrating a reduction in maximum localization error from 0.58 m to 0.20 m and mean error from 0.35 m to 0.12 m.

\end{abstract}

\renewcommand{\IEEEkeywordsname}{Index Terms}

\section{Introduction}



\begin{figure}
    \centering
    \includegraphics[width=3.3in]{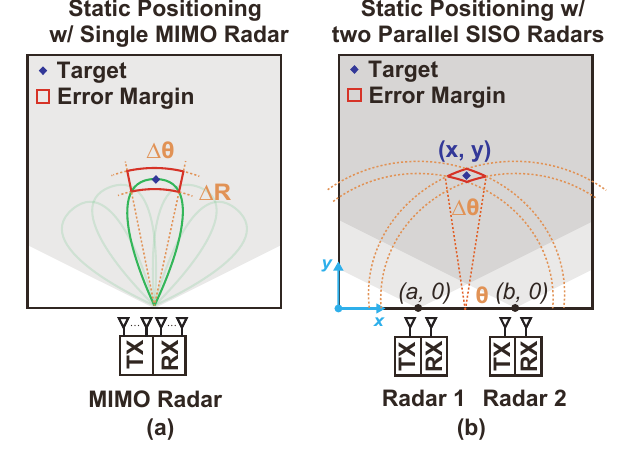}
    \caption{ Schematic comparison of localization error margins. (a) A single MIMO radar. (b) The proposed multi-site SISO configuration.
    }
    \label{fig:main}
\end{figure}

\IEEEPARstart{M}{imo} radar positioning is widely used in indoor sensing and monitoring. While commercial millimeter-wave radars have achieved centimeter-scale range resolution, their localization accuracy is fundamentally constrained by the angular resolution $\Delta\theta$ of the physical aperture as shown in Fig.~\ref{fig:main}(a). This geometric bottleneck is particularly detrimental in cluttered indoor environments, where multipath effects and angular glint severely degrade the phase-sensitive angle-of-arrival (AoA) estimation essential for compact monostatic multiple-input multiple-output (MIMO) systems.

Multi-site radar systems have emerged as a compelling research frontier due to their superior localization precision compared to conventional monostatic radar systems as shown in Fig.~\ref{fig:main}(b)\cite{qin_Indoor_2024}. To overcome these physical bottlenecks, distributed architectures leverage spatial diversity. For such multi-site systems, localization uncertainty is typically evaluated using the geometric dilution of precision (GDOP)\cite{aubry2023robust}. However, GDOP remains a dimensionless geometric multiplier, disconnecting the spatial topology from the physical SNR constraints defined by the Cramér-Rao lower bound (CRLB)\cite{2025_TMTT_multipath}. Consequently, a unified theoretical framework to benchmark low-cost, multi-site parallel single-input single-output (SISO) radars against compact monostatic MIMO systems is still missing. Although~\cite{Buchberger_APMC2022} introduced the concept of equivalent angular resolution, their model is restricted to targets along the system's central axis, failing to capture performance differences across the entire field of view (FoV).


In this paper, we propose a rigorous theoretical framework to evaluate the localization error margin in indoor positioning systems. We explicitly derive the equivalent angular resolution for SISO configurations, bridging the comparative gap between distributed and centralized topologies. This metric projects the geometric advantage of the distributed baseline into the angular domain, providing a unified basis for comparison against MIMO beam widths. Extensive simulations and experiments with 60-GHz commercial radars demonstrate that the proposed multi-site SISO configuration significantly outperforms monostatic MIMO, reducing the maximum localization error from 0.58 m to 0.20 m and the mean localization error from 0.35 m to 0.12 m in a $4\text{m} \times 4\text{m}$ environment.

\section{Theory}

\subsection{Signal Processing Model}
Consider an FMCW radar system where the beat signal, $s_{if}(t)$, resulting from mixing the transmitted and received signals, is expressed as
\begin{equation}
    s_{if}(t) = \frac{A}{2} \cos(2\pi k t_d t + \phi_0)
    \label{eq:beat_signal}
\end{equation}
where $k=B/\tau$ is the chirp slope, $t_d = 2r/c$ is the round-trip delay, and $r$ denotes the target range. Upon performing a discrete Fourier transform (DFT) and applying a constant false alarm rate (CFAR) detector, the range estimates for the distributed nodes, $r_1$ and $r_2$, are extracted from the range bin indices $m_{1,2}$
\begin{equation}
    r_{1,2} = \frac{c m_{1,2}}{2B}
    \label{eq:range_est}
\end{equation}
These post-detection range measurements serve as the primary observables for the subsequent multilateration process.

\subsection{Unified Error Analysis Framework}

\subsubsection{Monostatic MIMO Radar Baseline}
For a conventional monostatic MIMO radar, the cross-range localization uncertainty, $\epsilon$ is fundamentally governed by the Rayleigh angular resolution $\Delta\theta$. The error margin is typically approximated as
\begin{equation}
    \epsilon \approx \frac{1}{2} r \sin(\Delta\theta),
    \label{eq:mimo_error}
\end{equation}
where $r$ is the distance between target and radar. The angular resolution $\Delta\theta$ is determined by the physical aperture length $L$
\begin{equation}
    \Delta\theta \approx \frac{\lambda}{L \cos\theta_s},
    \label{eq:mimo_res}
\end{equation}
where $\theta_s$ is the steering angle. This dependence on $L$ creates the geometric bottleneck depicted in Fig. \ref{fig:main}(a).

\subsubsection{Multi-site SISO Radar Baseline}
The multi-site SISO radar systems localize the target primarily using post-detection range measurements. The post-detection range between the target and the two radars $r_1$ and $r_2$ can be obtained by~(\ref{eq:range_est}). 
For dual parallel radar systems as Fig.~\ref{fig:main}(b) shows, the target's coordinates $(x,y)$ meet the equation
\begin{equation}\label{eq:geometric mode}
\left\{
\begin{array}{l}
(x - a)^2 + y^2 = r_1^2, \\
(x - b)^2 + y^2 = r_2^2,
\end{array}
\right.
\end{equation}
where $a$ and $b$ denote the positions of the radar sensors in the room coordinate system. Then the target's coordinates can be solved by triangulation of the distance measurements as
\begin{equation}
\left\{
\begin{array}{l}
x = |\dfrac{a^{2} - b^{2} - r_{1}^{2} + r_{2}^{2}}{2(a - b)} |, \\
y = \sqrt{r_{1}^{2} - (x - a)^{2}}.
\end{array}
\right.
\end{equation}

Therefore, the distance and azimuth angle from point $(x,y)$ to $\left( \dfrac{a+b}{2},0 \right)$ is
\begin{equation}
\left\{
\begin{array}{l}
r_{eq} = \dfrac{\sqrt{2(r_{1}^{2} + r_{2}^{2}) - (a - b)^{2}}}{2}, \\
\theta = \arctan\left(\dfrac{y}{\dfrac{a+b}{2} - x}\right),
\end{array}
\right.
\end{equation}

where azimuth angle $\theta$ is between the position vector and the positive x-axis. Therefore, we describe the target position as the range measurement's function $\theta=f(r_{1},r_{2})$. The equivalent angular resolution can be described as
\begin{align}
\Delta \theta_{eq} &= \theta_{1} - \theta_{2} \notag \\
&\approx f(r_{1},r_{2}+\Delta r) - f(r_{1}+\Delta r,r_{2})
\label{eq:equivalent angle resolution}
\end{align}
where $\theta_{1}$ and $\theta_{2}$ are the azimuth angle corresponding to the boundary point. This formulation allows, for the first time, a direct comparison between the ``virtual aperture'' formed by distributed nodes and the physical aperture of MIMO arrays. When $x \approx (a+b)/{2}$ we can prove the formula is equivalent to ~\cite{Buchberger_APMC2022}. 

Therefore, localization error margin of multi-site radar positioning systems is
\setcounter{equation}{9} 
\begin{equation}
    \epsilon' \approx \dfrac{1}{2} r_{eq} \sin(\Delta \theta_{eq}). 
    \label{eq:localization error}
\end{equation}

From an information-theoretic perspective, $\Delta\theta_{eq}$ is not just a geometric parameter but a direct manifestation of the Fisher information gain provided by spatial diversity. In monostatic MIMO systems, the CRLB for angular variance is inversely proportional to the physical aperture size squared ($\text{CRLB}_{\theta} \propto (L \cdot \sqrt{\text{SNR}})^{-1}$)\cite{2025_TMTT_multipath}. In our distributed SISO architecture, the baseline separation acts as a synthesized virtual aperture. By minimizing $\Delta\theta_{eq}$, we are effectively maximizing the effective aperture in the Fisher information matrix, thereby reducing the CRLB. Thus, the proposed equivalent angular resolution serves as a unified bridge, translating the spatial geometry of distributed nodes into an equivalent SNR gain comparable to monostatic physical arrays.

\section{Simulation And Experiments}
\begin{figure}[t] 
    \centering
    \includegraphics[width=3.2in]{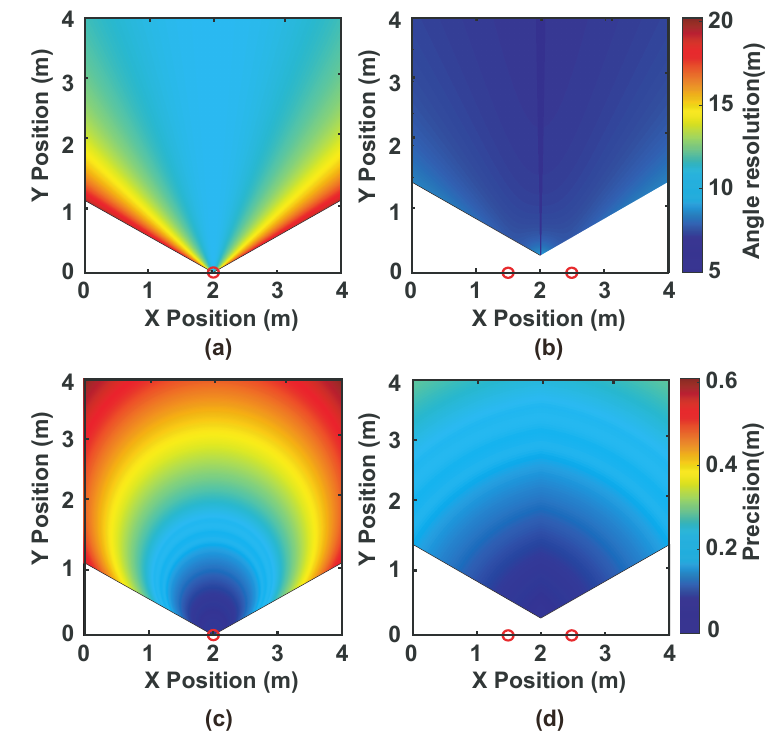}
    \caption{Simulation results comparison. (a) Angular resolution distribution of a single MIMO radar. (b) Equivalent angular resolution distribution of the proposed multi-site SISO system. (c) and (d) are spatial localization error map for the single MIMO radar and the multi-site systems.
    }
    \label{fig:precision}
\end{figure}

To validate the proposed theory, simulations were conducted with the parameter settings: $ B = 1 $\,GHz, $ N_{channel} = 8 $, $M_{sampling} = 256 $ and $f_c = 60 $ \,GHz.
Fig.~\ref{fig:precision}(a) shows the single MIMO radar angular resolution distribution. The radar is placed at $(2,0)$\,m. The angular resolution is mainly decided by the angle. Fig.~\ref{fig:precision}(b) shows the multi-site radar systems equivalent angular resolution. The dual radars are placed at $(1.5,0)$\,m and $(2.5,0)$\,m. The equivalent angular resolution is related to both the equivalent angle and the equivalent range. However, its variation is not as significant as in MIMO radar systems. Moreover, in the given space, the equivalent angular resolution of multi-site radar systems is always less than single MIMO radar. 

Fig.~\ref{fig:precision}(c) and (d) shows spatial localization error in single MIMO radar and multi-site SISO radar systems. The error is calculated by~(\ref{eq:localization error}). For MIMO radar systems, the angular resolution significantly increases with increasing angle. Consequently, areas at the same distance from the radar but with larger angles experience greater localization error.
In contrast, for multi-site radar systems, the effect of angle variation on the equivalent angular resolution is not significant. Therefore, the localization error in these systems primarily depends on the equivalent range, and multi-site radar positioning systems still have less error in most of the space.
\begin{figure}[t] 
    \centering
    \includegraphics[width=3.2in]{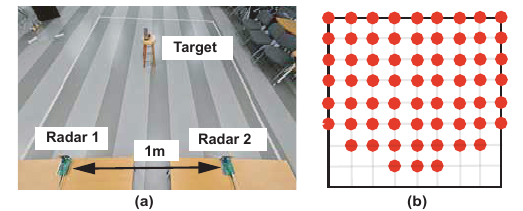}
    \caption{ Experimental setup and environment. (a) Photograph of the conference room showing multipath scattering conditions. (b) Schematic of the measurement grid.
    }
    \label{fig:setup}
\end{figure}
\begin{figure}[t] 
    \centering
    \includegraphics[width=3.4in]{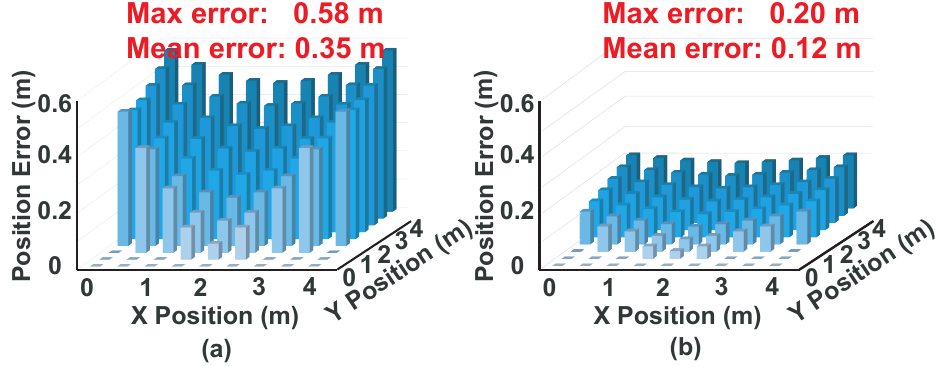}
    \caption{Experimental results comparison. (a) Positioning error in single MIMO radar. (b) Positioning error in the proposed multi-site system.
    }
    \label{fig:final}
\end{figure}

Experiments were further conducted to validate the theory and simulations. We have set up the experiment in the conference room and marked $4$\,m $\times$ $4$\,m space for measurements as shown in Fig.~\ref{fig:setup}(a). This operational environment presents significant multipath challenges. Conventional monostatic MIMO systems heavily rely on accurate AoA estimation, which is susceptible to angular glint and multipath distortions in such cluttered scenarios. In contrast, the proposed multi-site configuration leverages spatial diversity. By cross-referencing range-only measurements from spatially separated viewpoints, the system inherently mitigates the impact of angular estimation errors caused by multipath effects, resulting in more robust localization performance. The radar systems are based on the off-the-shelf commercial TI AWR6843ISK millimeter wave radar, the operating mode can switch between MIMO mode and SISO mode by controlling the antenna numbers. Data acquisition across the distributed nodes is performed independently, with temporal alignment achieved via host timestamps. Since the target is static, the millisecond-level synchronization error is acceptable. 
The radar units are positioned at coordinates $(1.5,0)$\,m and $(2.5,0)$\,m which are consistent with the aforementioned simulations. 

Considering the field of view~(FoV) of radar is $120^\circ$, we just choose red point position as measurement position during each measurement as shown in Fig.~\ref{fig:setup}(b), every red point is \SI{0.5}{m} apart from each other. The radar device parameter is the same as the simulation before. In the first test, we used single MIMO radar placed at $(2,0)$\,m. In the second test, the two radars were spaced at $(1.5,0)$\,m and $(2.5,0)$\,m. Precise spatial calibration is critical for minimizing systematic errors in the triangulation process. The baseline distance between the two radar centers was calibrated using a laser rangefinder with millimeter-level accuracy. Furthermore, although the proposed system utilizes non-coherent processing (relaxing the requirement for strict carrier phase synchronization), the spatial coordinates of the sensors were strictly aligned to the established coordinate system to ensure the validity of the geometric model in (\ref{eq:geometric mode}).


We repeated the experiment $3$ times and used the average of the $3$ measurements as the error value for each position. The results are shown in the Fig.~\ref{fig:final}. The localization error exhibits a similar error distribution to the simulation results in Fig.~\ref{fig:precision}(c) and (d). The maximum localization error margin is \SI{0.58}{m} in single MIMO radar systems and \SI{0.20}{m} in multi-site radar positioning systems, and the mean localization error margin is \SI{0.35}{m} and \SI{0.12}{m}.

\section{Conclusion}
This paper provides a comprehensive theoretical and experimental investigation into the localization error margin of multi-site radar positioning systems. The proposed framework provides a generalized closed-form analytical solution for equivalent angular resolution and distance metrics tailored for parallel SISO radar configurations, allowing the derivation of the localization error margin in a manner comparable to monostatic MIMO radar systems. This system holds significant promise for applications such as indoor elderly vital sign monitoring and fall detection. 

\section*{Acknowledgment}
This work is supported by the Shenzhen Science and Technology Program under Grant (JCYJ20230807091814030, JCYJ20220818100408018 and 20231115204236001), the National Natural Science Foundation of China under Grant (62471211 and 32371992), and in part by the Guangdong Basic and Applied Basic Research Foundation under Grant (2025A1515011109 and 2024A1515011902).
\bibliographystyle{IEEEtran}
\bibliography{lumped}

\vfill


\end{document}